\documentstyle[twoside,fleqn,fps,espcrc2]{article}

\newcommand{\AmS}{{\protect\the\textfont2
  A\kern-.1667em\lower.5ex\hbox{M}\kern-.125emS}}

\def\lsi{\raise0.3ex\hbox{$<$\kern-0.75em\raise-1.1ex\hbox{$\sim$}}}
\def\gsi{\raise0.3ex\hbox{$>$\kern-0.75em\raise-1.1ex\hbox{$\sim$}}}

\newcommand{\gsim}{\mathop{\gsi}}
\newcommand{\R}{{\kern+.25em\sf{R}\kern-.78em\sf{I} 
  \kern+.78em\kern-.25em}}
\newcommand{\C}{{\kern+.25em\sf{C}\kern-.50em\sf{I} \kern+.50em\kern-.25em}}
\newcommand{\eps}{\epsilon}
\newcommand{\be}{\begin{equation}}
\newcommand{\ee}{\end{equation}}
\newcommand{\bea}{\begin{eqnarray}}
\newcommand{\eea}{\end{eqnarray}}
\newcommand{\nn}{\nonumber}
\newcommand{\la}{\langle}
\newcommand{\ra}{\rangle}
\newcommand{\vnu}{\vert \nu \vert}
\newcommand{\figsizeA}{\epsfxsize=48mm}

\title{Meson Correlation Functions in the
 $\eps$-Regime 
\thanks{Talk presented by T. Chiarappa at Lattice 2003
\newline \hspace*{0.5mm} Preprint DESY 03-146, HU-EP-03/60, SFB/CPP-03-37
}}
\author{T. Chiarappa
\address{ NIC/DESY Zeuthen, Platanenallee 6, D-15738 Zeuthen, Germany \\
$^{{\rm b}}$ Institut f\"{u}r Physik, Humboldt Universit\"{a}t zu Berlin,
Newtonstr. 15, D-12489 Berlin, Germany }
, W. Bietenholz $^{{\rm b}}$, K. Jansen $^{{\rm a}}$,
K.-I. Nagai $^{{\rm a}}$ and S. Shcheredin $^{{\rm b}}$  
}

\begin{document}

\begin{abstract}

We present a numerical pilot study of the meson correlation functions
in the $\eps$-regime of chiral perturbation theory ($\chi$PT).
Based on simulations with overlap fermions we
measured the axial and pseudo-scalar correlation functions,
and we discuss the implications for the leading low energy
constants in the chiral Lagrangian.

\vspace*{-4mm}

\end{abstract}

\maketitle

\section{THEORETICAL BACKGROUND}

Ginsparg-Wilson fermions \cite{GW,Has} obey an exact, lattice modified
chiral symmetry \cite{ML}. Therefore they have exact zero modes,
which provides together with the Index Theorem a conceptually clean
definition of the topological charge \cite{Has}.

In particular the overlap operator \cite{HN} represents a relatively
simple solution to the Ginsparg-Wilson relation. It allows us to
tackle now a number of physically interesting issues,
which were inaccessible to numerical studies before.
In particular, there is hope for simulations
to penetrate the regime of light mesons, which is described
by $\chi$PT. The latter involves low energy constants as free
parameters, which can in principle be determined by QCD simulations
in that regime. 
To the lowest order, the effective chiral Lagrangian 
of $\chi$PT takes the form
\bea
{\cal L}_{\rm eff}[U] &=& 
\frac{F_{\pi}^{2}}{4} {\rm Tr} \, \left[ \partial_{\mu} U
\partial_{\mu} U \right]  \\
& - & \frac{\Sigma}{2} {\rm Tr} \left[ 
{\cal M} (e^{i\theta /N_{f}} U + e^{-i\theta /N_{f}} 
U^{\dagger} ) \right]  \nn
\eea
where $U(x) \in SU(N_{f})$, 
$N_f$ is the number of flavors,
${\cal M}$ is the (diagonal) quark mass matrix
and $\theta$ the vacuum angle. 
The pion decay constant $F_{\pi}$ and the scalar condensate $\Sigma$
appear here as the leading low energy constants.

From the practical point of view, the $\eps$-regime of $\chi$PT
looks particularly attractive because it deals with small volumes
\cite{GasLeu}.
This regime is characterized by the hierarchy relation
\be
\Lambda_{\rm QCD}^{-1} \ll L \ll m_{\pi}^{-1}
\ee
so that the pion correlation length $m_{\pi}^{-1}$ is much larger than 
the linear size of the system $L$. Moreover, $\Lambda_{\rm QCD}$ 
can be related to the cutoff of the effective theory,
$\Lambda_{\rm QCD} \sim 4 \pi F_{\pi}$.
One identifies the matrix field as
$U = \exp ( i \sqrt{2} \xi /F_{\pi})$, where the field
$\xi (x)$ describes the light mesons.
The $\eps$-expansion then relies on a non-perturbative
treatment of the zero-mode, whereas the excitations 
(the non-zero modes of $\xi (x)$) do fit into the
volume and can be evaluated perturbatively.
In the $\eps$-regime, the topological charge $\nu$ of the gauge field
plays an important r\^{o}le \cite{LeuSmi}; observables should be
measured at fixed values of $\vnu$. Note that the values of $F_{\pi}$
and $\Sigma$ in the $\eps$-regime are the same as in the physical
situation of an infinite volume.

Hence the main virtues of Ginsparg-Wilson fermions --- no additive
mass renormalization and a clean definition of the topological charge ---
render this lattice fermion formulation ideally suited for
simulations in the $\eps$-regime.
However, they
are numerically so demanding that for the time being only quenched
QCD simulations are feasible. We performed such simulations with
overlap fermions and
the Wilson gauge action at $\beta =6$ on lattices of size
$10^3 \times 24$ and $12^4$. The mass of the Wilson kernel
in the overlap operator was set to $-1.4$, which is optimal
for locality \cite{HJL}.

\begin{figure}[hbt]
\vspace*{-8mm}
\def\fpsangle{270}
\figsizeA
\fpsbox{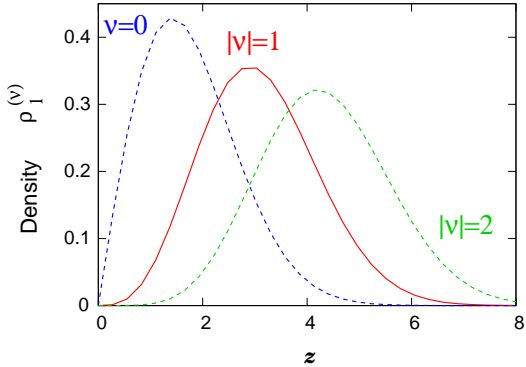}
\vspace*{-9mm}
\caption{\it{The probability distribution for the first non-zero
EV of the massless Dirac operator, according to Random Matrix Theory.}}
\label{RMT-EV}
\vspace*{-8mm}
\end{figure}

Let us mention that Random Matrix Theory applied to
QCD yields predictions for the low lying eigenvalues (EVs) $i \lambda$
of the Dirac operator \cite{RMT}. Simulations in the $\eps$-regime
confirm these predictions, at least for the first non-zero
EV, if the physical volume is not too small, $L \gsim
1.2 ~ {\rm fm}$ \cite{BJS}.\footnote{A lower bound for $L$ 
has also been obtained from the condition
for the $\eps$ parameter to be sufficiently small \cite{PO}.}
Fig.\ \ref{RMT-EV} shows the probability distribution $\rho_{1}^{(\nu )}$
of this first EV in various topological sectors, as a function of the
dimensionless variable ${\bf z}= \lambda \Sigma V$.
In particular we see that in the neutral sector there is a significant
probability $\rho_{1}^{(0)}$
for very small EVs. In turned out that this property
makes measurements very hard, i.e.\ they would require
a huge statistics \cite{nuzero}. 
However, the situation is clearly better at
non-trivial topology. Therefore we concentrate here on the
sector $\vnu =1$.

We measured various types of {\em meson correlation functions}
at momentum $\vec p = \vec 0$ and Euclidean time $t$.
The corresponding formulae predicted by quenched $\chi$PT 
in a volume $L^3 \times T$ are
given in Refs.\ \cite{corre}. They were obtained consistently from
two methods denoted as ``replica'' \cite{replica} and ``supersymmetric'' 
\cite{SUSY}. The vector correlation function
vanishes identically.
The simplest non-trivial case is the bare axial correlation 
function,
\vspace*{-1mm}
\bea  \label{AA}
\hspace*{-7mm} && \la A_{\mu}(t) \, A_{\mu}(0) \ra_{\nu} = 
\frac{F_{\pi}^{2}}{T} 
+ 2 m \, \Sigma_{\vnu}(z) \, T \cdot
h_{1}(\tau ) \\
\hspace*{-7mm} &&
2 h_{1}(\tau ) = \tau^{2} - \tau + 1/6 
\ , \ \tau = t/T \ , \ z = m \Sigma V \ , \nn \\
\hspace*{-7mm} &&
\frac{\Sigma_{\nu}(z)}{\Sigma} = z \Big[ I_{\nu}(z) K_{\nu}(z) + 
I_{\nu +1}(z) K_{\nu -1}(z) \Big] + \frac{\nu}{z} \nn ,
\eea
where $I_{\nu}$ and $K_{\nu}$ are modified Bessel functions
and $m$ is the bare quark mass.

To the first order the pseudo-scalar correlation function reads
\bea
\hspace*{-7mm} &&
\la P(t) \, P(0) \ra_{\nu} = \frac{\Sigma_{\vnu}^{{\rm 1-loop}}(z)}
{2 m T} L^{3} - \frac{\Sigma^{2}}{2F_{\pi}^{2}} h_{1}(\tau ) \times \nn \\
\hspace*{-7mm} &&
\Big[ \frac{\alpha}{3} c_{+}^{(\vnu )}(z) - b_{+}^{(\nu )}(z) \Big] 
- \frac{\Sigma^{2}}{6 F_{\pi}^{2}} h_{2}(\tau ) m_{0}^{2} 
c_{+}^{(\vnu )}(z) \ , \nn \\
\hspace*{-7mm} &&
h_{2}(\tau ) = - \frac{T^{3}}{24} \Big( \tau^{4} - 2\tau^{3} 
+ \tau^{2} - \frac{1}{30} \Big) \ , \nn \\
\hspace*{-7mm} &&
\Sigma_{\nu}^{{\rm 1-loop}}(z) = \Sigma_{\nu}(z) \frac{z'}{z} \ ,
\quad z' = m \, \Sigma_{\rm eff}(V) \, V \ ,  \label{PsPs}
\eea
In this case, and also in the scalar correlation function, 
new parameters of the effective chiral Lagrangian have to be
taken into account, which are specific for the quenched approximation:
$m_{0}$ and $\alpha$ are a scalar mass and a kinetic 
coupling, respectively. For details and for 
the form of $\Sigma_{\rm eff}$,
$c_{+}^{(\nu )}$ and $b_{+}^{(\nu )}(z)$
we refer to Refs.\ \cite{corre}.


\section{NUMERICAL RESULTS}

We first consider the axial correlation function because it only 
involves the constants $\Sigma $ and $F_{\pi}$. On the $10^3 \times 24$
lattice the spatial extent is below $1~{\rm fm}$, and our data show
clearly that in such a small volume $\chi$PT is not applicable,
as expected.

Therefore we now present results from the $12^4$ lattice, which corresponds
to a volume of $(1.12~{\rm fm})^{4}$, i.e.\ close to the point where
the applicability of $\chi$PT should set in. 
Our quark mass amounts to $m = 21.3~ {\rm MeV}$,
and our statistics involves $78$ configurations with $\nu = \pm 1$.

\begin{figure}[hbt]
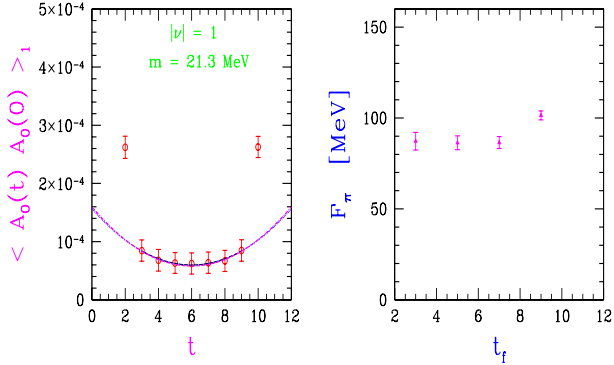

\def\fpsangle{270}
\epsfxsize=48mm
\fpsbox{newfit.epsi}
\vspace*{-10mm}
\caption{\it{The axial correlation function (left) and the
values of $F_{\pi}$ obtained from fits over $t_{f}$ points around the 
center $t=6$ (right).}}
\label{axiFig}
\vspace*{-9mm}
\end{figure}

Fig.\ \ref{axiFig} (on the left) shows our data for the axial correlator.
The curve corresponding to
eq.\ (\ref{AA}) can be fitted well over some interval that excludes
the points near the boundary. This fixes $F_{\pi}$, which enters in an 
additive constant. On the right of Fig.\ \ref{axiFig}
we plot the resulting $F_{\pi}$
as a function of the number $t_f$ of $t$ values (around $T/2$) that
we include in the fit. We find a decent plateau, which suggests
--- up to renormalization --- a value of
\be
F_{\pi} = (86.7 \pm 4.0) ~ {\rm MeV} \ .
\ee
On the other hand, $\Sigma$ cannot be easily extracted from these
fits. It is related to the curvature in the minimum, but if we vary
$\Sigma$ over a wide range the fit function is hardly affected.
In fact, Fig. \ref{axiFig} (left) shows fits to $t_f = 7$ points
for $\Sigma$ varying from $0$ to $(250~{\rm MeV})^{3}$, but those curves
can hardly be distinguished.

\begin{figure}[hbt]
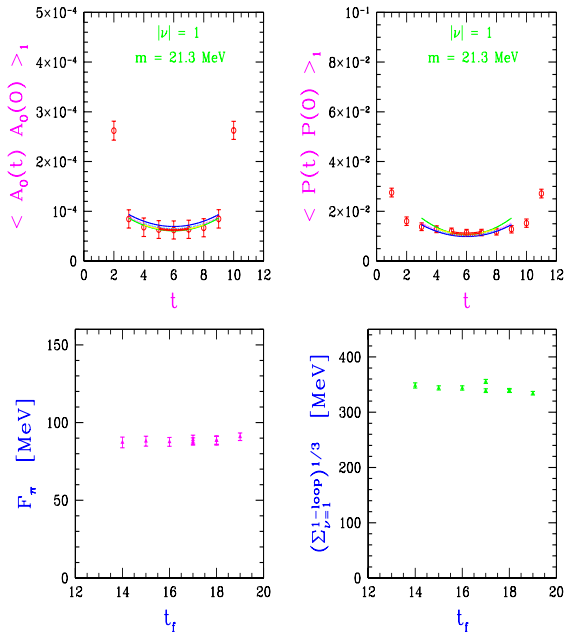

\vspace*{-12mm}
\def\fpsangle{0}
\epsfxsize=74mm
\fpsbox{ax_ps2.epsi}
\vspace*{-10mm}
\caption{\it{Above: measured data and fits for
the axial and pseudo-scalar correlation functions.
Below: The values $F_{\pi}$ and $\Sigma_{1}^{{\rm 1-loop}}$ 
obtained from five parameter fits.
Here $t_{f}$ is the sum of points considered in the simultaneous 
fits for both correlation functions.}}
\label{5fit}
\vspace*{-10mm}
\end{figure}

As an attempt to proceed also in that respect, we show in Fig.\
\ref{5fit} (above) the results for the axial and pseudo-scalar correlation
functions, which are fitted simultaneously.
Here the fits involve the parameters $\alpha$, $m_0$
and $\Sigma_{1}^{{\rm 1-loop}}$ (introduced in eq.\ (\ref{PsPs})), 
in addition to $F_{\pi}$ and $\Sigma$.
These five parameter fits lead to decent plateaux simultaneously for
$F_{\pi}$ and $\Sigma_{1}^{{\rm 1-loop}}$, as Fig.\ \ref{5fit}
(below) illustrates. $\Sigma_{1}^{{\rm 1-loop}}$
depends on the other four parameters, and it represents 
a one loop approximation to $\Sigma$.

\section{CONCLUSIONS}

\vspace*{-2mm}

We presented results of a pilot study of meson correlation functions
in the $\eps$-regime, based on quenched QCD with overlap fermions.
To avoid trouble with very low EVs we worked in the sector
$\vnu =1$.
Although the statistics is still rather modest, we recognize
some qualitative features.
If the physical volume is sufficiently large, the data
can be fitted to the functions predicted by quenched $\chi$PT.
These fits allow for a good determination of $F_{\pi}$, whereas
the evaluation of $\Sigma$ is far more difficult.


\vspace*{-2mm}

\begin{thebibliography}{40}

\vspace*{-1mm}

\bibitem{GW} P.\ Ginsparg and K.\ Wilson, Phys.\ Rev.\ D25 (1982) 2649.

\bibitem{Has} P.\ Hasenfratz, V.\ Laliena and F.\ Niedermayer,
Phys.\ Lett.\ B427 (1998) 317.

\bibitem{ML} M.\ L\"uscher, Phys.\ Lett.\ B428 (1998) 342.

\bibitem{HN} H.\ Neuberger, Phys.\ Lett.\ B417 (1998) 141.

\bibitem{GasLeu} J.\ Gasser and H.\ Leutwyler, Phys.\ Lett.\ B188 (1987) 477.

\bibitem{LeuSmi} H.\ Leutwyler and A.\ Smilga, Phys.\ Rev.\ D46 (1992) 5607.

\bibitem{HJL} P.\ Hern\'{a}ndez, K.\ Jansen and M.\ L\"uscher,
Nucl.\ Phys.\ B552 (1999) 363.

\bibitem{RMT} P.H.\ Damgaard and S.M.\ Nishigaki, 
Nucl.\ Phys.\ B518 (1998) 495.
T.\ Wilke, T.\ Guhr and T.\ Wettig, Phys.\ Rev.\ D57 (1998) 6486.

\bibitem{BJS} W.\ Bietenholz, K.\ Jansen and S.\ Shcheredin, 
JHEP 07 (2003) 033. See  also contributions by
S.\ Shcheredin ({\tt hep-lat/0309030}), 
T.\ Streuer and P.\ Weisz to these proceedings.

\bibitem{PO} S.\ Prelovsek and K.\ Orginos, Nucl. Phys. B
(Proc. Suppl.) 119 (2003) 822.

\bibitem{nuzero} P.\ Hern\'{a}ndez, K.\ Jansen and L.\ Lellouch,
Phys.\ Lett.\ B469 (1999) 198.
K.-I. Nagai, these proceedings ({\tt hep-lat/0309051}).

\bibitem{corre} P.H.\ Damgaard, M.C.\ Diamantini, P.\ Hern\'{a}ndez
and K.\ Jansen, Nucl.\ Phys.\ B629 (2002) 226.
P.H.\ Damgaard, P.\ Hern\'{a}ndez, K.\ Jansen, M.\ Laine and L.\ Lellouch,
Nucl.\ Phys.\ B656 (2003) 226.

\bibitem{replica} P.H.\ Damgaard and K.\ Splittorff, 
Phys. Rev. D62 (2000) 54509.

\bibitem{SUSY} A.\ Morel, J.\ Physique 48 (1987) 1111.
C.\ Bernard and M.\ Golterman, Phys.\ Rev.\ D46 (1992) 853.

\end{thebibliography}
\end{document}